\documentclass[aps,prl,twocolumn,floatfix,nofootinbib,showpacs]{revtex4}
\usepackage{graphicx,color}
\begin{document}

\title{Partially Strong $WW$ Scattering}
\renewcommand{\thefootnote}{\fnsymbol{footnote}}

\author{
Kingman Cheung$^{1,2}$, Cheng-Wei Chiang$^{3,4}$, Tzu-Chiang Yuan$^2$}
\affiliation{
$^1$ Department of Physics, National Tsing Hua University,
Hsinchu, Taiwan \\
$^2$
Physics Division, National Center for Theoretical Sciences, Hsinchu, Taiwan\\
$^3$ Department of Physics and Center for Mathematics and Theoretical
Physics, National Central University, Chung-Li, Taiwan \\
$^4$ Institute of Physics, Academia Sinica, Taipei, Taiwan 
}
\renewcommand{\thefootnote}{\arabic{footnote}}
\date{\today}

\begin{abstract}
%  We consider the scenario in which only a light Higgs boson is
%  discovered at the CERN LHC.  
  What if only a light Higgs boson is discovered at the CERN LHC?
  Conventional wisdom tells us that the
  scattering of longitudinal weak gauge bosons would not grow strong
  at high energies. We show
  that this is not always true.  In some composite models,
  two-Higgs-doublet models, or even supersymmetric models, the
  presence of a light Higgs boson does not guarantee the complete
  unitarization of the $WW$ scattering.
  After the partial unitarization by
  the light Higgs boson, the $WW$ scattering becomes strongly
  interacting until it hits one or more heavier Higgs bosons or other
  strong dynamics.  We analyze how the LHC experiments
  can reveal this interesting possibility of partially strong $WW$
  scattering.
\end{abstract}
\maketitle

{\it Introduction --}  
The CERN Large Hadron Collider (LHC) will commence soon to uncover the
mystery of electroweak symmetry breaking (EWSB).  The ultimate
goal of the LHC is to search for the Higgs boson and hopefully any new
physics beyond the standard model (SM).  Physicists have been exciting
about mapping new observations to the parameter spaces in various
models, known as the ``inverse LHC problem.''
However, one may anticipate that only one light Higgs boson is found
in the first few years of LHC run.  This is perhaps one of the most
pessimistic scenarios.  A light Higgs boson $h$ of mass $m_h \alt
130$ GeV can be discovered through the $\gamma\gamma$ or $b\bar b$
modes.  Since this mass is below the $WW$ or $ZZ$ threshold, it would
be hard to probe how much this light Higgs boson is directly linked to
EWSB. Several recent works have suggested
precision measurements in the branching ratios of the light Higgs boson
\cite{Giudice,wise,Randall}  and $W_LW_L$ scattering 
\cite{Giudice,Randall} to unravel the nature of EWSB.

In this Letter, we propose to use the scattering of longitudinal weak
gauge bosons to probe whether the light Higgs boson completely or just
partially unitarizes the scattering amplitudes.  Longitudinal
weak gauge boson scattering is an old idea \cite{WW} and it has been
used to impose a unitarity bound on the mass of a heavy Higgs boson.  
At high
energies, the longitudinal components of the weak gauge bosons recall
their identities as the Goldstone bosons of the EWSB sector \cite{equiv}.
The scattering amplitudes of these Goldstone bosons with purely gauge
contributions grow with energy as $s/m_W^2$, where $s$ is the
squared center-of-mass (CM) energy of the $W_LW_L$ system.  Here we used
$W$ to generically denote either a $W$ or $Z$ boson, unless 
otherwise stated. In the SM
with a light Higgs boson, the amplitude will be completely unitarized
by the Higgs boson. % and the scattering processes remain perturbative.
Once $\sqrt s$ goes beyond the light Higgs boson mass, the
scattering amplitude will no longer grow like $s/m_W^2$.
% but remains a constant.  
%%
If the SM
with a light Higgs by itself were indeed an ultraviolet (UV) 
complete theory, that
would be our final prediction albeit a boring one.  However, many
issues such as the fine-tuning problem in the Higgs boson mass,
nonzero neutrino masses, dominant dark matter and dark energy contents
in the Universe do not have easy solutions within the SM if not
impossible.  They all suggest that new physics must be involved in
order to solve some or all of these puzzles.

In many extensions of the SM, {\it e.g.}, two-Higgs-doublet model
(2HDM), little Higgs model, etc, there is usually one light Higgs
boson.  However, the light Higgs boson may not be fully responsible
for the symmetry breaking, so that longitudinal $W_LW_L$
scattering is only partially unitarized by the light Higgs boson.
Such an idea was recently mentioned first in Ref. \cite{Giudice}
and then in Ref. \cite{Randall}.
Terms effectively scaling like $s/m_W^2$ in the scattering amplitude
comes back such that it becomes strong after hitting the light
Higgs pole.  
%At the LHC, this will give rise to observable signals \cite{madison}.
%%
At a sufficiently high energy, there will be the other part of EWSB
sector, {\it e.g.}, the heavier Higgs boson of the 2HDM or the UV
completion of the little Higgs models, to eventually unitarize the $W_LW_L$
scattering.
Nonetheless, if the scale of this UV part is far enough from the light
Higgs boson, 
%the unitarization mechanism is not efficient enough and
the onset of strong $W_LW_L$ scattering between the light Higgs mass
and the UV scale should be discernible at the LHC.  
The main result of this work shows that longitudinal weak gauge
boson scattering can indeed provide a useful means to probe the 
nature of EWSB associated with a light Higgs boson.

{\it Methodology --}
In the SM, the $hWW$ coupling is $g^{\rm SM}_{hWW} g^{\mu\nu} \equiv g
m_W g^{\mu\nu}$, where $g$ is the $SU(2)$ gauge coupling constant.
For a concrete
example consider the scattering of $W^+_L W^-_L \to W^+_L W^-_L$,
which proceeds through the $t$- and $s$-channels of $\gamma$ and $Z$
exchanges, the 4-point vertex, and the $s$- and $t$-channels of 
Higgs exchanges. The longitudinal polarization of
the $W$ boson can be expressed as $\epsilon^\mu_L(p) = p^\mu / m_W +
v^\mu(p)$ with $v^\mu(p) \simeq -m_W /(2 {p^0}^2) ( p^0, - \vec{p})
\sim O(m_W/E_W)$.  In the CM system of $W^+_L (p_1) W^-_L
(p_2) \to W^+_L(k_1) W^-_L(k_2)$, one can choose $v^\mu(p_1) = -
2 (m_W/s ) p_2^\mu$, and so on.  The sum of the amplitudes of all 
gauge diagrams is, in the high energy limit,
\begin{equation}
\label{amp-gauge}
i {\cal M}^{\rm gauge} = - i \frac{g^2}{4 m_W^2}\, u + O\left((E/m_W)^0\right)
 \;,
\end{equation}
where $E$ denotes the scattering energy.  Note that the quartic term
proportional to $E^4 /m_W^4$ naively expected from the 4-point vertex
is canceled by the $\gamma$- and $Z$-exchange diagrams.
%so that the gauge amplitude grows only quadratically as $E^2/m_W^2$ in
%Eq.~(\ref{amp-gauge}).  
On the other hand, the sum of the two Higgs diagrams is
\begin{eqnarray}
\label{amp-Higgs}
i {\cal M}^{\rm Higgs} &=& - i \frac{g^2}{4 m_W^2} \left[
  \frac{ (s - 2 m_W^2)^2}{s -m_h^2} + \frac{(t - 2 m_W^2)^2}{t -m_h^2} 
  \right ] \nonumber \\
 & \simeq &
  i \frac{g^2}{4 m_W^2} \, u  \;,
\end{eqnarray}
in the limit of $s \gg m_h^2, m_W^2$.  Thus, the bad energy-growing
term is delicately canceled between the gauge diagrams and the Higgs
diagrams.  This is a well-known fact in the SM.  However, in some
extended models that the light Higgs boson has only a fraction of the
SM coupling strength with the gauge bosons, one expects the gauge
amplitude to keep growing with $s$ after hitting the light Higgs
pole.

\begin{figure}[t!]
\centering
\includegraphics[width=3.3in]{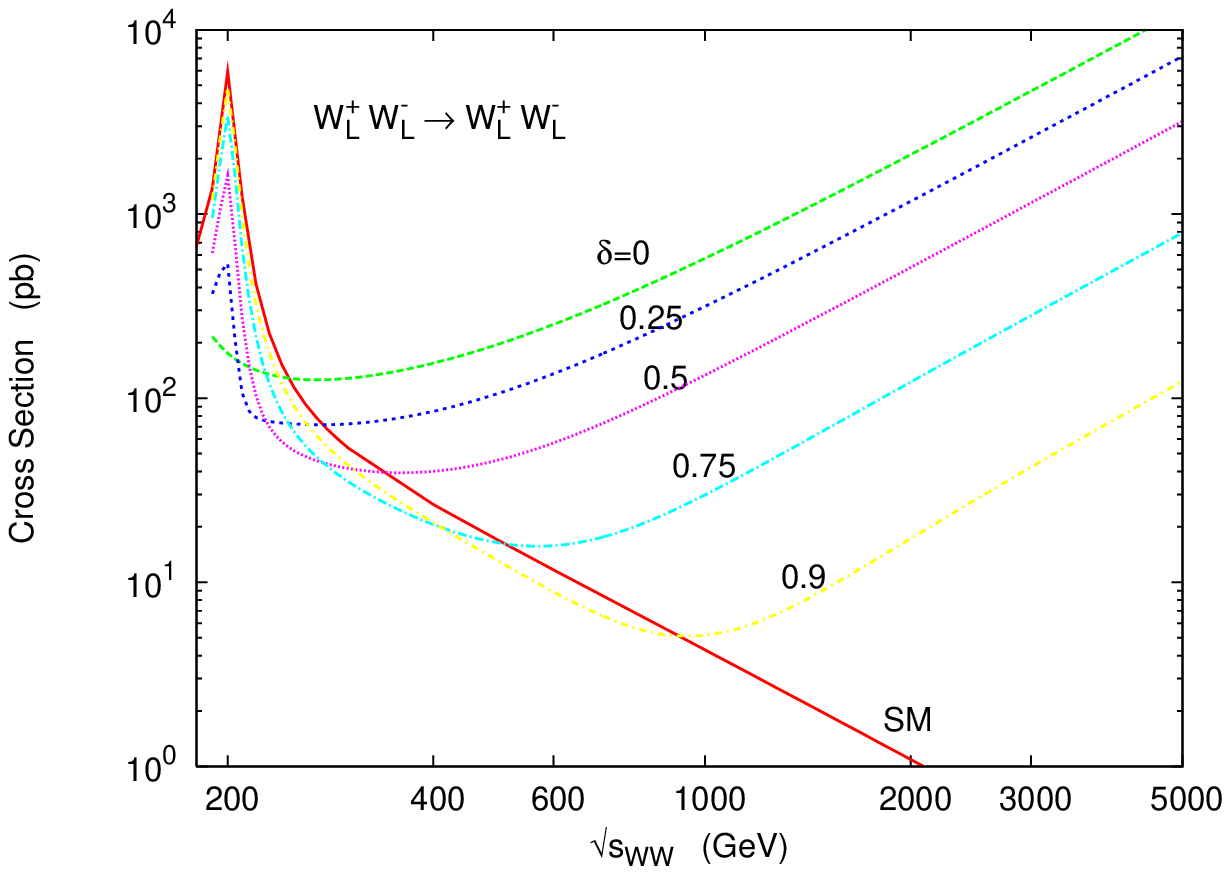} \\
(a) \\
\includegraphics[width=3.3in]{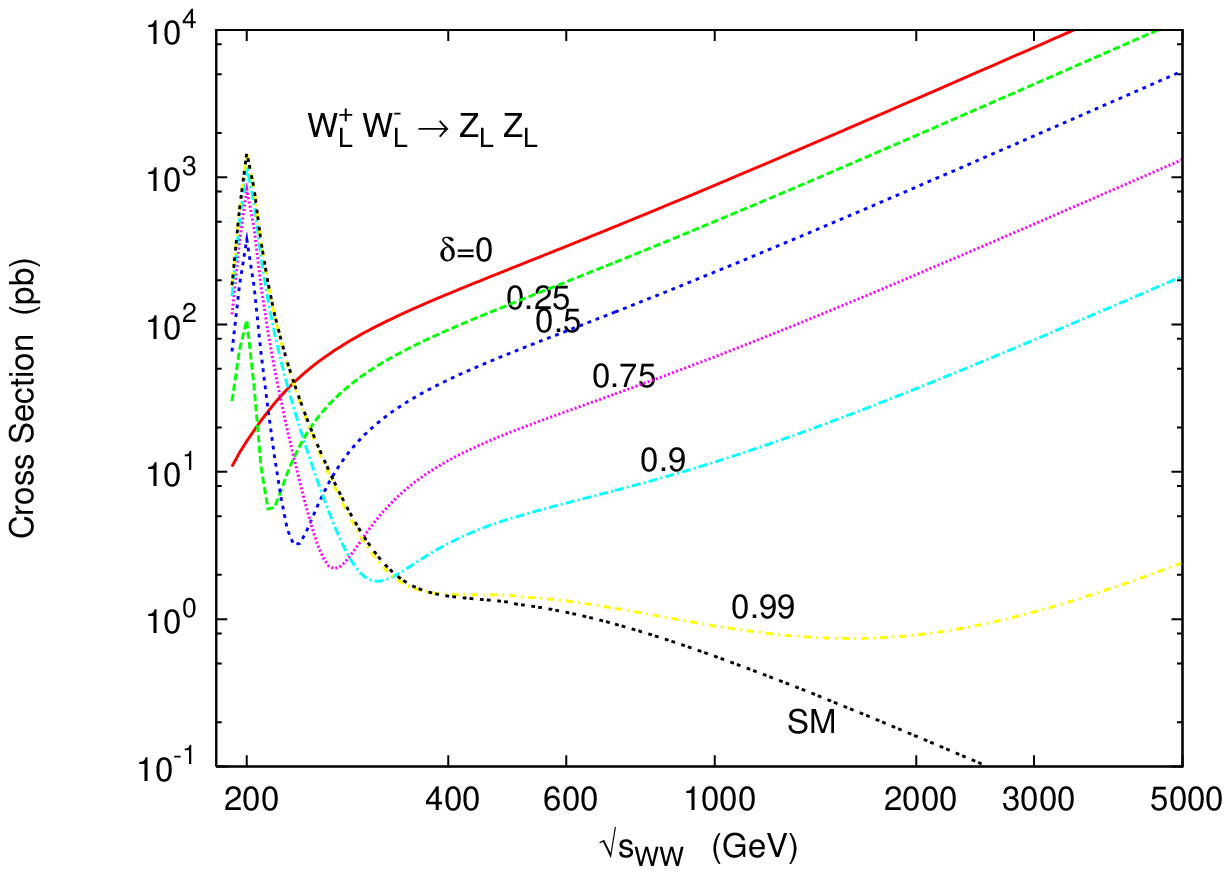} \\
(b) \\
\caption{\label{ww} \small Scattering cross sections for (a) $W^+_L
  W^-_L \to W^+_L W^-_L$ and (b) $W^+_L W^-_L \to Z_L Z_L$ versus
  $\sqrt{s_{WW}}$.  Various values of $\delta$ are shown, where
  $\delta$ denotes the size of the Higgs amplitude relative to the SM
  one.  An angular cut of $|\cos \theta_{WW}|< 0.8$ is applied and the
  light Higgs boson mass $m_h = 200$ GeV is assumed.}
\end{figure}

Given our ignorance of what may lie beyond the SM, we follow the
approach adopted by recent studies \cite{Giudice,wise,Randall} to
parametrize the coupling $g_{hWW}$ as a fraction $\sqrt{\delta}$ of
its SM value.\footnote{ We note that in models with an extra $Z'$
  boson, it is possible to have the $hZZ$ coupling modified due to
  $Z-Z'$ mixing while the $hWW$ coupling remains intact. The
  unitarization of the longitudinal weak gauge boson scattering in
  such models is somewhat different from what we discuss in this work
  and deserves a separate study.}  As a result, the Higgs
amplitude in Eq.~(\ref{amp-Higgs}) becomes $\delta$ times the SM
value.  For small enough $\delta$, the total scattering amplitude will
grow after the light Higgs pole due to incomplete cancellation of the
bad high-energy behavior terms.  This is true even for a rather large
$\delta=0.9$.  We show in Fig.~\ref{ww}(a) the exact scattering cross
sections for $W^+_L W^-_L \to W^+_L W^-_L$ versus $\sqrt{s_{WW}}$,
where we have assumed $m_h=200$ GeV.  For the SM case the sum of
amplitudes converges to $O((E/m_W)^0)$ terms, and the cross section
thus drops like $1/s_{WW}$.  When the size of the Higgs amplitude
deviates from the SM value, even with a small amount (say
$\delta=0.9$), the cross section will cease falling but start climbing
instead around $\sqrt{s_{WW}} \alt 1$ TeV.  It turns around at lower
$\sqrt{s_{WW}}$ for smaller $\delta$'s.  A similar behavior happens in
the $W^+_L W^-_L \to Z_L Z_L$ channel, as shown in Fig.~\ref{ww}(b),
where the turn-around occurs at even lower energies.
Not so dramatic feature can also be shown for the nonresonant
channels, such as $W^\pm_L W^\pm_L \to W^\pm_L W^\pm_L$ and $W^\pm_L
Z_L \to W^\pm_L Z_L$, where the cross sections only climb up
gradually.
Such behavior can be readily observed at the LHC.  We will give some
realistic event numbers later to support our claim.

\begin{figure}[t!]
\centering
\includegraphics[width=3.3in]{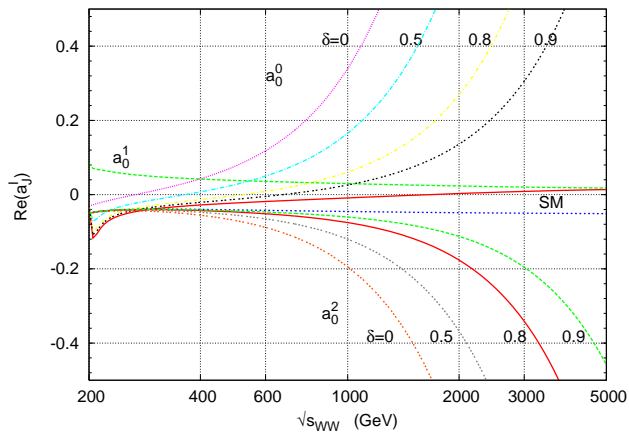}
\caption{\small \label{limit} The partial-wave coefficients
  $a^{0,1,2}_0$ versus the center-of-mass energy $\sqrt{s_{WW}}$ for
  various $\delta=0-0.9$.  }
\end{figure}

We also analyze the partial-wave coefficients to determine when the
unitarity is violated.  Consider a clean isospin $I=2$ channel, $W^+_L
W^+_L \to W^+_L W^+_L$.
The sum of the gauge amplitudes
$
i {\cal M}^{\rm gauge} = i \frac{g^2}{4} 
  \left[ \frac{u + t}{m_W^2}  +O\left( (E/m_W)^0 \right )\right],
$
in which the quartic terms proportional to $E^4/m_W^4$ are again canceled.
The SM Higgs boson with a full strength $g^{\rm SM}_{hWW}$ contributes
$
i {\cal M}^{\rm Higgs} = - i \frac{g^2}{4} \left[
  \frac{u+t}{m_W^2} + O\left( (E/m_W)^0 \right) \right ],
$
which is valid for $|t|,|u| \gg m_W^2, m_h^2$.  
It is clear that the bad energy-growing terms
cancel each other such
that their sum behaves well at high energies.
Now as before, we assume that the coupling $g_{hWW}$ is a fraction
$\sqrt{\delta}$ of its SM value so that the cancellation is only
partial.
In the high energy limit, the total amplitude becomes
\begin{equation}
\label{simple}
i {\cal M}^{\rm gauge} + i {\cal M}^{\rm Higgs} \simeq 
i \frac{g^2}{4 m_W^2} (u+t) \, ( 1-\delta) \;.
\end{equation}
One can then check the unitarity limit as a function of $\delta$.  The
partial-wave coefficient for the dominant $S$-wave scattering is
\begin{equation}
a^2_0 = \frac{1}{64\pi} \int^1_{-1} d \cos\theta \, {\cal M}\left (
W^+_L W^+_L \to W^+_L W^+_L \right )
\;,
\end{equation}
where the superscript 2 denotes the isospin of the $W^+_LW^+_L$
system.  Besides this $I=2$ channel, one can also study
the $I = 0, 1$ partial waves.  Unitarity requires $|\Re e\; a^I_0| \le
1/2$.
%
%\begin{eqnarray}
%a^0_0 &=& \frac{1}{64\pi} \int^1_{-1} d \cos\theta \, \left[
%3\, {\cal M}\left ( W^+_L W^-_L \to Z_L Z_L \right ) \right. \nonumber \\
%& + & \left. {\cal M} \left(W^+_L W^+_L \to W^+_L W^+_L \right ) \right ] \; ,
%\end{eqnarray}
%
%\begin{eqnarray}
%a^1_0 & = & \frac{1}{64\pi} \int^1_{-1} d \cos\theta \, \left\{
% 2  \left[ {\cal M}\left(W^+_L W^-_L \to W^+_L W^-_L \right ) \right. \right. \nonumber \\
%& & \quad\quad\quad\quad\quad\quad\quad\quad
% -  \; {\cal M} \left. \left ( W^+_L W^-_L \to Z_L Z_L \right ) \right]
%\nonumber \\
%& - & {\cal M} \left. \left( W^+_L W^+_L \to W^+_L W^+_L \right ) \right\}  \; .
%\end{eqnarray}
%
%
We show in Fig.~\ref{limit} the partial-wave coefficients $a^I_0$ $(I
= 0,1,2)$ versus $\sqrt{s_{WW}}$ for various $\delta =0-0.9$.  We use
the full expressions of the amplitudes, instead of the simplified
expression like Eq.~(\ref{simple}), in the evaluations.  Details of
these amplitudes will be presented elsewhere.  
At high energies,
$a^0_0$'s are positive while $a^2_0$'s stay negative.  The unitarity
limit can be read off when each curve reaches $\Re e(a^I_0) = \pm
1/2$.  
%%%
 Note that the matrix element of the $I=1$ channel at high energy
is an odd function of $\cos\theta$  such that the 
partial wave $a^1_0$  does not show any growing behavior for 
various $\delta$. 
The unitarity limits that would be obtained from $a^1_1$
are significantly weaker than those from $a^{0,2}_0$
due to $P$-wave suppression.  The most severe
violation of unitarity is in the $a^0_0$ channel.  For example,
unitarity is violated at $\sqrt{s_{WW}}\simeq 1.7 \; (2.7)$ TeV for
$\delta = 0.5\;(0.8)$.  The LHC may not be able to directly probe such
high CM energies.  But the growing behavior of the
scattering amplitudes should be discernible at much lower energies.
%The $W^+_L W^+_L$ is a non-resonant channel.  Although the presence of
%the light Higgs boson is not manifest, the amount of rising of 
%the amplitude with $s$ can determine the size of $g_{hWW}$ relative to the
%SM value.
%

%Therefore, in the experiment one can measure the partial wave coefficient
%versus the invariant mass of the $WW$ system, so that one can tell if
%the light Higgs boson can completely cancel the unitarity growing.

{\it Various models --} The simplest example of partially strong weak
gauge boson scattering is the 2HDM \cite{Randall}, in which light
Higgs boson couples to the vector boson with a strength $g_{hWW} =
\sin (\beta - \alpha) g_{hWW}^{\rm SM}$, where $\tan\beta$ is the
ratio of the VEVs of the two doublets and $\alpha$ is the mixing angle
of the two CP even neutral Higgs bosons.  If the other neutral Higgs
boson $H$ is much heavier, the weak gauge boson scattering
amplitudes will enjoy their growths as $s/m_W^2$ for the energy between
the two Higgs boson masses.
This heavier neutral Higgs boson
couples to the weak gauge boson with a strength $g_{HWW} = \cos
(\beta-\alpha) g_{HWW}^{\rm SM}$ such that it can unitarize the rest
of the growing amplitudes when $s_{WW} > m_H^2$.  A general 2HDM
has enough room in the parameter space to allow
$\sin(\beta-\alpha)$ to be small while keeping the other Higgs boson $H$
heavy.
However, in minimal supersymmetric standard models (MSSM) 
the heavier the heavy Higgs boson $H$ is, the closer to $1$ the factor 
$\sin(\beta-\alpha)$ will be.
As shown in Ref. \cite{CP}, it is possible to achieve a light Higgs boson
with a small $\sin(\beta-\alpha)$ while keeping the other neutral ones
relatively light as well. Thus, no appreciable strong weak gauge boson scattering 
can be observed in the MSSM.

%ii) 
In the strongly-interacting light Higgs model \cite{Giudice}, 
a composite-like model for the light Higgs boson is assumed 
with the size of the ratio $g_{hWW} / g_{hWW}^{\rm SM} $ smaller than 1.  
All other
heavier degrees of freedom are integrated out and the effects are
parameterized as an effective Lagrangian with an explicit UV cutoff.  
The partial widths of the light Higgs boson will be affected.
Also, the weak gauge boson scattering amplitudes described by some higher 
dimensional 
effective operators  will also grow with $s$ until the cutoff is reached.
Similarly, in a model of multi-scalar doublets \cite{wise} all the heavy
Higgs bosons can be integrated out to give corrections to the partial decay
widths of a light Higgs boson, which will affect significantly its 
discovery modes at the LHC.

\begin{figure}[t!]
\centering
\includegraphics[width=3.3in]{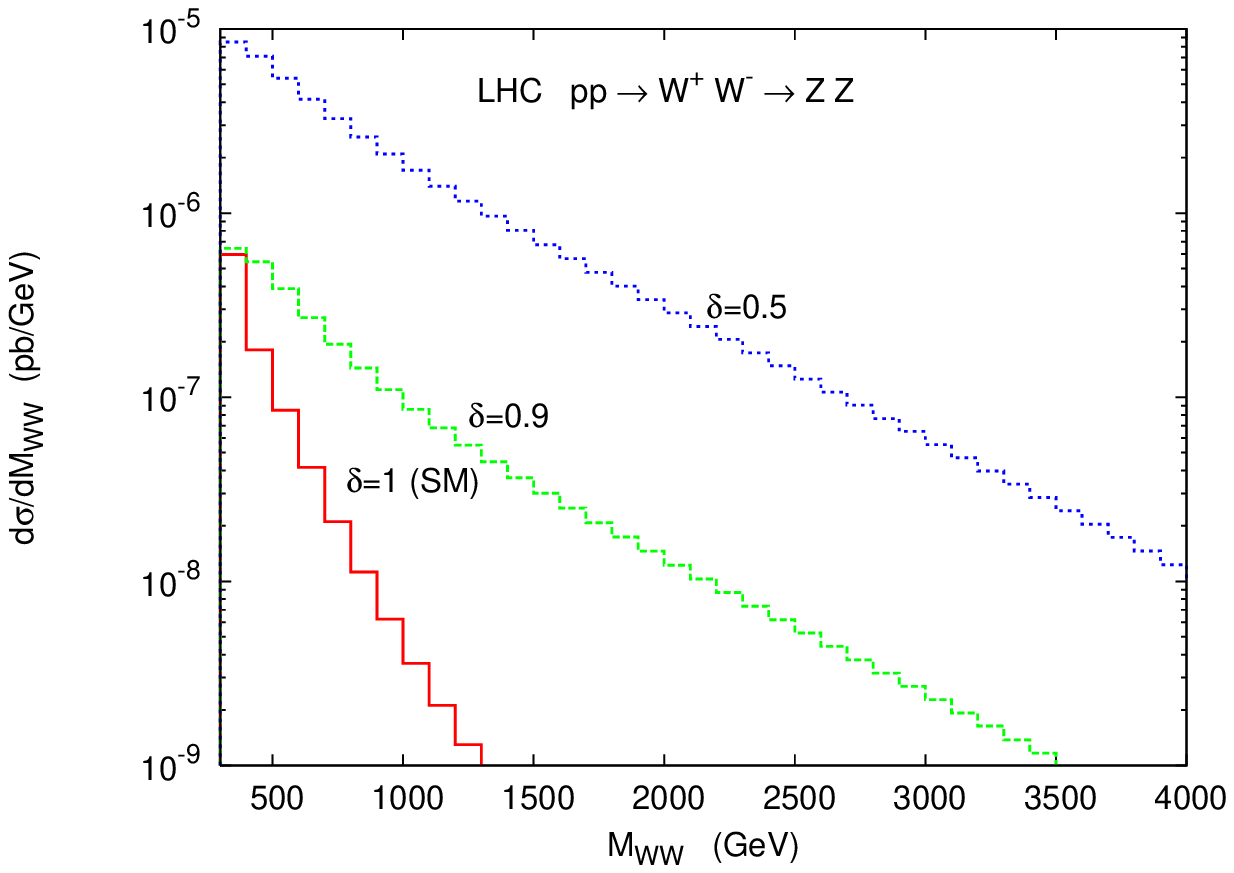}
\includegraphics[width=3.3in]{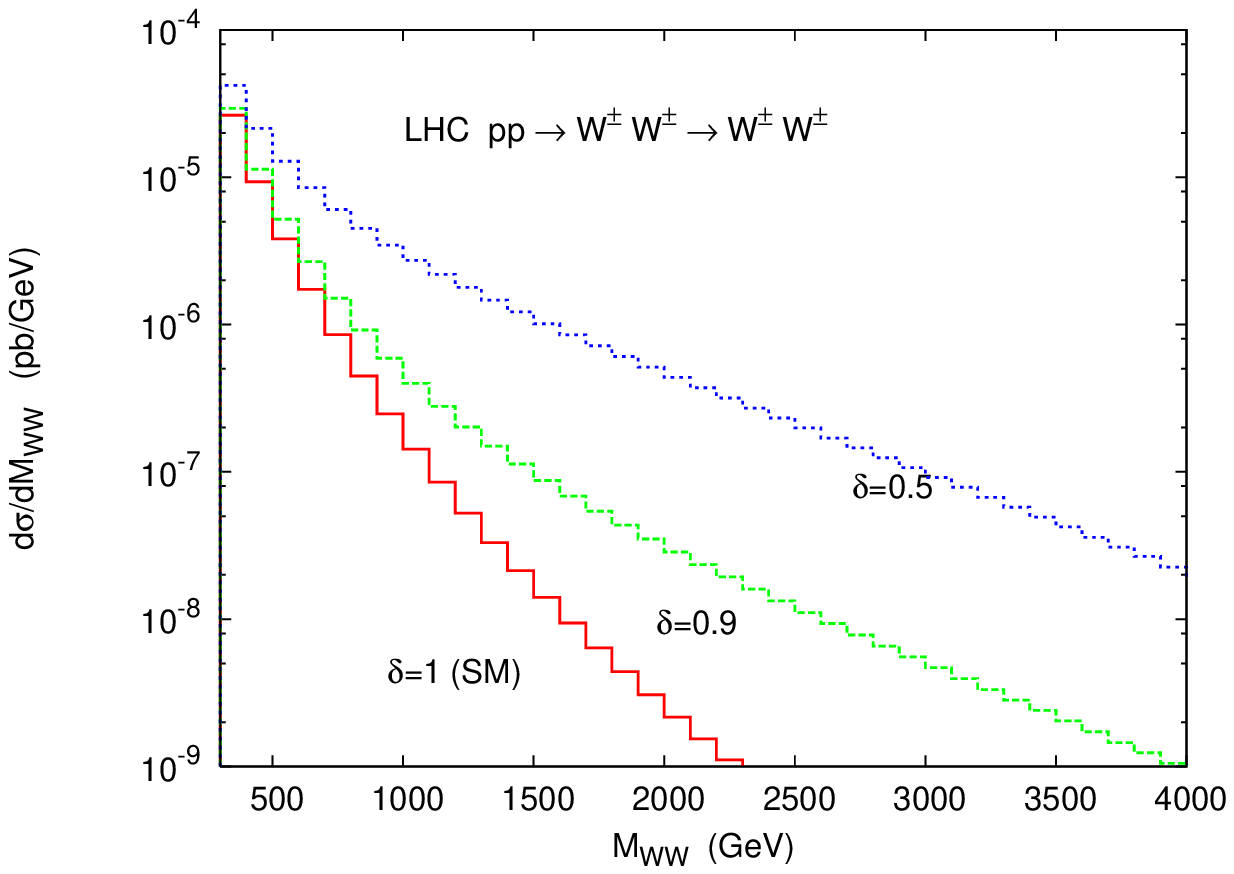}
\caption{\small \label{ewa-ww}
Invariant mass distribution for 
(a) $p p \to W^+_L W^-_L X \to Z_L Z_L$,
(b) $p p \to W^{\pm}_L W^{\pm}_L X \to W^{\pm}_L W^{\pm}_L $
for $\delta = 1, 0.9$ and $0.5$ at the LHC using EWA and $m_h = 200$ GeV.
}
\end{figure}

{\it LHC signals --} We show the invariant mass spectrum in 
Fig. \ref{ewa-ww} for $pp \to W^+_L W^-_L \to Z_L Z_L$ and 
$pp \to W^\pm_L  W^\pm_L \to W^\pm_L W^\pm_L$.
In Fig. \ref{ewa-ww}(a), the mere $\delta=0.9$ curve is above the SM one
for $M_{WW}> 300$ GeV, in accord with Fig.~\ref{ww}(b), while $\delta=0.5$ 
case is way above the SM prediction.  The nonresonant channel $W^\pm_L W^\pm_L$
shown in Fig. \ref{ewa-ww}(b) requires a smaller $\delta$ in order to see a 
large deviation from the SM. 
\begin{table*}[th!]
\caption{Event rates for longitudinal weak gauge boson
scattering at the LHC with a yearly 
luminosity of 100 fb$^{-1}$ using the EWA for 
$\delta = 1$ (SM), $0.9$, $0.5$ and 0 (No Higgs).  Branching ratios 
for the leptonic 
final states are summed for $\ell = e$ and $\mu$. We set 
$m_h = 200$ GeV and $M_{WW}^{\rm min} = 300$ GeV. 
\label{table}
}
\begin{ruledtabular}
\begin{tabular}{l|cccc}
$\quad\quad\quad\quad$Subprocess &  & Number of Events &\\
& $\delta$ = 1 (SM) & 0.9 & 0.5 & 0 (No Higgs)\\
\hline
$W^{\pm}_L W^{\pm}_L \to  W^{\pm}_L W^{\pm}_L \to \ell^\pm \nu \ell^\pm  \nu$  
      & 21 & 26 & 57 & 118  \\ 
\hline
$W^{\pm}_L W^{\mp}_L \to  W^{\pm}_L W^{\mp}_L \to \ell^\pm \nu \ell^\mp \nu$  
      & 8 & 7 & 17 &  67 \\ 
\hline
$W^{\pm}_L Z_L \to  W^{\pm}_L Z_L \to \ell^\pm \nu \ell^+\ell^- $  
   & 4  & 5  & 13 & 33 \\ 
\hline
$W^{+}_L W^{-}_L \to  Z_L Z_L \to \ell^+\ell^- \ell^+ \ell^-$  
    & 0.04  & 0.12 & 2 & 9 \\ 
$W^{+}_L W^{-}_L \to  Z_L Z_L \to \ell^+\ell^-  \nu \bar \nu$  
   & 0.25  & 0.74 & 12 & 50 \\ 
$Z_L Z_L \to  Z_L Z_L \to \ell^+ \ell^- \ell^+ \ell^-$  & 0.4 & 0.32 & 0.08 & 0 \\ 
$Z_L Z_L \to  Z_L Z_L \to \ell^+ \ell^- \nu \bar \nu$  & 2.4 & 2 & 0.5 & 0\\ 
\end{tabular}
\end{ruledtabular}
\end{table*}
We mainly focus on leptonic final states, $WW \to \ell\nu\ell\nu$,
$ZZ \to \ell^+ \ell^-\ell^+ \ell^-$ and $ZZ \to \ell^+ \ell^-\nu\bar \nu$.
The latter mode is used because the four charged-lepton mode of $ZZ$ 
is too small
for realistic event rates.  We show the event rates at the LHC for 
various scattering channels in Table \ref{table}, with an angular cut
of $|\cos\theta_{WW}|< 0.8$ and $M_{WW} > 300$ GeV.  
We use the naive effective $W$-boson approximation (EWA) \cite{ewa} to
estimate the event rates, which is good enough to demonstrate
the main idea here.
The studies of strongly-interacting weak gauge boson scattering and
various backgrounds were summarized in Refs. \cite{madison}, based on
the techniques of central-jet vetoing and forward jet-tagging.  The
jet-tagging and central-jet vetoing efficiencies under optimized cuts
were listed there too.  The event rates predicted in this work are to be 
multiplied by those efficiencies.
It is easy to see that with $\delta =0.5$ significant enhancement to the 
event rates relative to the SM can be realized.

%{\it Conclusion --} 
To conclude, detailed studies of longitudinal weak gauge boson
scattering at the LHC can provide useful hints of new physics at a
higher scale, despite only a light Higgs boson 
may be discovered during the
first few years at the LHC.  If unitarity is only partially
fulfilled by the light Higgs, the scattering cross sections must be
growing as energy increases before it reaches the other heavier Higgs
bosons or other UV completions to achieve the full unitarization. These
partial growths of the cross sections can be discernible at the LHC 
provided that the UV part is at a high scale. This can be realized in
two- or multi-Higgs-doublet models with large $\tan\beta$ as was
studied recently in Refs.~\cite{wise,Randall}, which proposed using the
precision measurements of light Higgs boson decays to explore effects
from new physics. Our approach of using longitudinal weak gauge
boson scattering is complementary to those works but more
direct. Discovery of a light Higgs together with positive
observations of partially strong $WW$ scattering at the
LHC will definitely indicate that the SM is just an effective theory of some
more fundamental theories.  
%Perhaps LHC will just reveal the tip of
%the iceberg.

{\it Acknowledgments.}
KC would like to thank the Institute of Theoretical Physics at the 
Chinese University of Hong Kong for hospitality.
This work was supported in part by the NSC of Taiwan and by the NCTS.

%%%%%%%%%%%%%%%%%%%%%%%%%%%%%%%%%%%%%%%%


\begin{thebibliography}{99}

\bibitem{Giudice}
  G.~F.~Giudice, C.~Grojean, A.~Pomarol and R.~Rattazzi,
  %``The Strongly-Interacting Light Higgs,''
  JHEP {\bf 0706}, 045 (2007)
  [arXiv:hep-ph/0703164].
  %%CITATION = JHEPA,0706,045;%%

\bibitem{wise}
S.~Mantry, M.~Trott and M.~B.~Wise,
  %``The Higgs Decay Width in Multi-Scalar Doublet Models,''
  Phys. Rev. D {\bf 77}, 013006 (2008),
  arXiv:0709.1505 [hep-ph].
  %%CITATION = ARXIV:0709.1505;%%

\bibitem{Randall}
  L.~Randall,
  %``Two Higgs Models for Large Tan Beta and Heavy Second Higgs,''
  JHEP {\bf 0802}, 084 (2008),
  arXiv:0711.4360 [hep-ph].
  %%CITATION = ARXIV:0711.4360;%%

\bibitem{WW}
B.~W.~Lee, C.~Quigg and H.~B.~Thacker,
  %``Weak Interactions At Very High-Energies: The Role Of The Higgs Boson
  %Mass,''
  Phys.\ Rev.\  D {\bf 16}, 1519 (1977);
  %%CITATION = PHRVA,D16,1519;%%
M.~S.~Chanowitz,
  %``The no-Higgs signal: Strong W W scattering at the LHC,''
  Czech.\ J.\ Phys.\  {\bf 55}, B45 (2005)
  [arXiv:hep-ph/0412203].
  %%CITATION = CZYPA,55,B45;%%

\bibitem{equiv}
Y.~P.~Yao and C.~P.~Yuan,
  %``Modification of the Equivalence Theorem Due to Loop Corrections,''
  Phys.\ Rev.\  D {\bf 38}, 2237 (1988);
  %%CITATION = PHRVA,D38,2237;%%
H.~G.~J.~Veltman,
  %``The Equivalence Theorem,''
  Phys.\ Rev.\  D {\bf 41}, 2294 (1990);
  %%CITATION = PHRVA,D41,2294;%%
H.~J.~He, Y.~P.~Kuang and X.~y.~Li,
  %``On the precise formulation of equivalence theorem,''
  Phys.\ Rev.\ Lett.\  {\bf 69}, 2619 (1992).
  %%CITATION = PRLTA,69,2619;%%

%\bibitem{CCY}
%K.~Cheung, C.W. Chiang and T.-C.~Yuan, in preparation.

\bibitem{CP}
M.~Drees,
  %``A supersymmetric explanation of the excess of Higgs-like events at LEP,''
  Phys.\ Rev.\  D {\bf 71}, 115006 (2005)
  [arXiv:hep-ph/0502075];
  %%CITATION = PHRVA,D71,115006;%%
A.~Belyaev {\it et al.},
  %``Light MSSM Higgs boson scenario and its test at hadron colliders,''
  arXiv:hep-ph/0609079; 
  %%CITATION = HEP-PH/0609079;%%
M.~Asano, S.~Matsumoto, M.~Senami and H.~Sugiyama,
  %``Neutralino Dark Matter in Light Higgs Boson Scenario,''
  arXiv:0711.3950 [hep-ph].
  %%CITATION = ARXIV:0711.3950;%%

\bibitem{ewa}
S.~Dawson,
  %``The Effective W Approximation,''
  Nucl.\ Phys.\  B {\bf 249}, 42 (1985).
  %%CITATION = NUPHA,B249,42;%%

\bibitem{madison}
J.~Bagger {\it et al.},
  %``The Strongly interacting W W system: Gold plated modes,''
  Phys.\ Rev.\  D {\bf 49}, 1246 (1994);
  %%CITATION = PHRVA,D49,1246;%%
  %``LHC analysis of the strongly interacting W W system: Gold plated modes,''
  {\it ibid.}  {\bf 52}, 3878 (1995).
  %%CITATION = PHRVA,D52,3878;%%

\end{thebibliography}
\end{document}